# Perpendicular magnetization reversal, magnetic anisotropy, multi-step spin switching, and domain nucleation and expansion in $Ga_{1-x}Mn_xAs$ films

X. Liu, W. L. Lim, L.V. Titova, M. Dobrowolska, and J. K. Furdyna

*Department of Physics, University of Notre Dame, Notre Dame, IN 46556*

M. Kutrowski and T. Wojtowicz

*Department of Physics, University of Notre Dame, Notre Dame, IN 46556*

*and*

*Institute of Physics, Polish Academy of Sciences, 02-668 Warsaw, Poland*

## Abstract

We present a comprehensive study of the reversal process of perpendicular magnetization in thin layers of the ferromagnetic semiconductor $Ga_{1-x}Mn_xAs$. For this investigation we have purposely chosen $Ga_{1-x}Mn_xAs$ with a low Mn concentration ($x \approx$ 0.02), since in such specimens contributions of cubic and uniaxial anisotropy parameters are comparable, allowing us to identify the role of *both* types of anisotropy in the magnetic reversal process. As a first step we have systematically mapped out the angular dependence of ferromagnetic resonance in thin $Ga_{1-x}Mn_xAs$ layers, which is a highly effective tool for obtaining the magnetic anisotropy parameters of the material. The process of perpendicular magnetization reversal was then studied by magneto-transport (i.e., Hall effect and planar Hall effect measurements). These measurements enable us to observe coherent spin rotation and non-coherent spin switching between the (100) and (010) planes. A model is proposed to explain the observed multi-step spin switching. The agreement of the model with experiment indicates that it can be reliably used for determining magnetic anisotropy parameters from magneto-transport data. An interesting characteristic of perpendicular magnetization reversal in $Ga_{1-x}Mn_xAs$ with low $x$ is the appearance of a double hysteresis loops in the magnetization data. This double-loop behavior can be understood by generalizing the proposed model to include the processes of domain nucleation and expansion.

# 1. INTRODUCTION

Ferromagnetic semiconductors (e.g., $Ga_{1-x}Mn_xAs$) have been the subject of detailed experimental and theoretical research for nearly a decade.[1,2,3,4,5,6,7] The ability of observing ferromagnetism in a semiconductor provided much of the early motivation for this activity.[8] Since that time considerable effort has focused on the complexities of exchange coupling,[9] the formation of domains,[10,11] and magnetic anisotropy[12,13,14] in order to understand the unusual magnetic properties that characterize these materials. For example, unlike ferromagnetic metals, in these new dilute magnetic systems the demagnetization field is insignificant, and magnetic crystalline anisotropy plays an essential role in determining magnetization reversal both for in-plane and for perpendicular configurations.[12] It is important to note that, since the magnetic properties of ferromagnetic semiconductors can be externally-controlled,[15,16] these materials are potential candidates for the design and implementation of magnetic logic devices – a feature that provides much of the motivation for the intense research activity currently under way in this area.

It is now well established within the framework of the Zener model of hole-mediated ferromagnetism[5,6] that the symmetry properties of the valence band are directly responsible for the magnetic anisotropy of $Ga_{1-x}Mn_xAs$; and that the Mn-ions favor the alignment of their magnetic moments either parallel or perpendicular to the plane, depending on whether the strain induced in $Ga_{1-x}Mn_xAs$ by the substrate on which it is grown is compressive or tensile.[17] In the case of compressive strain, the in-plane uniaxial and cubic anisotropies in thin III-Mn-V films have been extensively explored by ferromagnetic resonance (FMR)[18,19], magneto-optic Kerr effect (MOKE)[12] and magneto-transport[3,13]. An impressive degree of insight has been gained from even relatively simple models – specifically, the combination of the coherent rotation model of Stoner and Wohlfarth and of non-coherent spin switching – for explaining the in-plane magnetization hysteresis loops. In contrast to this, the perpendicular anisotropy in the same situation has thus far not been extensively explored. In particular, mechanisms of the perpendicular magnetization reversal have not yet been unambiguously identified.[12,20]

In this paper, we focus on four interrelated subjects: (1) *the mapping of magnetic anisotropy*; (2) *the process of perpendicular magnetization reversal*; (3) *multi-step spin switching*; and (4) *domain nucleation and expansion*. The mapping of magnetic anisotropies in the $Ga_{1-x}Mn_xAs$ film – including both perpendicular and in-plane uniaxial and cubic anisotropies – is obtained by fully mapping the angular dependence of ferromagnetic resonance (FMR) at various temperatures and its analysis.

For this investigation we have purposely chosen $Ga_{1-x}Mn_xAs$ with a low Mn concentration ($x \approx 0.02$), since in such specimens contributions of cubic and uniaxial anisotropy parameters are comparable, allowing us to identify the role of *both* types of anisotropy in the magnetic reversal process. The process of perpendicular magnetization reversal itself can be described using the coherent rotation model of Stoner and Wohlfarth, including thermal fluctuations. To obtain a picture of the multi-step spin switching which accompanies the magnetization reversal, we have used magneto-transport measurements that involve both the anomalous Hall effect and the planar Hall effect. This combination of the Stoner and Wohlfarth model and non-coherent spin switching has given very satisfactory results in reproducing the basic features of perpendicular magnetization reversal observed experimentally. Finally, we note that the multi-step profile of the magnetization reversal observed in the magneto-transport measurements is accompanied by the appearance of a striking double-hysteresis loop. We show that this last feature can be qualitatively understood in term of the nucleation and expansion of magnetic domains.

## 2.   EXPERIMENTAL DETAILS

The $Ga_{1-x}Mn_xAs$ samples for this study were grown on (001) semi-insulating "epi-ready" GaAs substrates in a Riber 32 R&D molecular beam epitaxy (MBE) system. The growth was monitored *in situ* by reflection high energy electron diffraction (RHEED). Standard effusion cells supplied the Ga and Mn fluxes, and the flux of $As_2$ was produced by a cracker cell. First a GaAs buffer of thickness 400 nm was grown at a temperature $T_S$ ~ 590°C (i.e., under standard GaAs growth conditions). The substrate was then cooled down to 210 °C for low-temperature (LT) growth. Using the $As_2$:Ga beam equivalent pressure ratio of 20:1, a 1 nm-thick buffer layer of LT-GaAs was then grown, followed by a $Ga_{1-x}Mn_xAs$ ($x < 0.03$) layer of 200 nm thickness. The RHEED pattern indicated (1×2) reconstruction during the low-temperature growth of $Ga_{1-x}Mn_xAs$, showing no evidence of precipitation of MnAs inclusions (i.e., no spotty RHEED pattern). Furthermore, clear RHEED oscillations indicative of two-dimension layer-by-layer growth were observed at the initial stages of $Ga_{1-x}Mn_xAs$ deposition. Monitoring the RHEED oscillations provided a precise measure of the growth rate of 0.26 ml/sec. The lattice constants of the resulting $Ga_{1-x}Mn_xAs$ layers were measured by x-ray diffraction (XRD) experiments using a double crystal diffractometer with Cu $K\alpha_1$ radiation. The Mn concentration $x$ was then determined from the lattice constant using the method of analysis developed by Schott *et al.* based on their detailed high-resolution XRD studies of on $Ga_{1-x}Mn_xAs$ mixed crystals over a wide range of Mn concentrations.[21]

Ferromagnetic resonance (FMR) measurements were carried out at 9.46 GHz using a Bruker electron paramagnetic resonance (EPR) spectrometer. In this article we will focus on the information obtained from the FMR position $H_R$ and its dependence on the orientation of the applied dc magnetic field **H** relative to the crystal axes of $Ga_{1-x}Mn_xAs$ specimens observed at different temperatures. For FMR measurements the MBE-grown $Ga_{1-x}Mn_xAs$ layer was cleaved into three square pieces with edges along the [110] and [$1\bar{1}0$] directions (see the coordinate system shown in Fig. 1). Each square piece was then cemented to a parallelepiped of the GaAs (100) SI substrate material (same material as that used for growth of the ferromagnetic film) in one of the three orientations shown in Fig. 1. Since the magnetic field of the EPR spectrometer is confined to the horizontal plane, this allowed us to carry out measurement on specimens with the layer plane mounted vertically in two basic configurations (referred to as Setup 1 and Setup 2). Specifically, taking the [$\bar{1}10$] edge of specimen to be vertical (Setup 1), this allowed measurement with the dc field **H** in any intermediate orientation between the normal to the layer plane, **H**‖[001], and the in-plane orientation **H**‖[110]. Similarly, when the [010] direction – i.e., the diagonal of square – is vertical (Setup 2), we could map out the FMR for field orientations between the normal orientation, **H**‖[001], and the in-plane orientation **H**‖[100]. Additionally, in the third configuration (Setup 3) the sample was mounted with the layer plane horizontal (i.e., the [001] direction pointing up). In this configuration we could measure the angular dependence of FMR when the field was confined to the layer [i.e., to the (001) plane], including the orientations $\mathbf{H}\|[110]$, $\mathbf{H}\|[1\bar{1}0]$ and $\mathbf{H}\|[100]$, as well as intermediate orientations in that plane.

The FMR studies were accompanied by magnetotransport (resistivity and Hall effect) measurements performed on samples cut from the same $Ga_{1-x}Mn_xAs$ layer in the six-probe Hall geometry with indium ohmic contacts; and by SQUID magnetometry characterization. Typically, Curie temperature of the $Ga_{1-x}Mn_xAs$ ($x = 0.02$) specimen is 40 K, and Hall effect measurements carried out at room temperature gave a hole concentration $1.1 \times 10^{20}$ cm$^{-3}$. Despite the complication due to the anomalous Hall effect, this result provides a reasonable estimate of the actual hole concentration in $Ga_{1-x}Mn_xAs$ samples with a low values of $x$.[22]

## 3.    RESULTS AND DISCUSSION

### 3.1.    Mapping magnetic anisotropy

It is now well established that the local Mn ions and the free holes in the III-Mn-V system form one "global" complex via strong magnetic coupling. [23] In the FMR experiment the total magnetic moment of this coupled complex precesses as a whole around the direction of the total static magnetic field present in the system (i.e., the sum of the applied magnetic field and magnetic anisotropy fields) at the Larmor frequency $\omega$.[24] When microwaves are present, absorption occurs when the microwave frequency coincides with the precession frequency. The precession of the magnetic moment can be described by the well-known Landau-Lifshitz-Gilbert equation.[25]

Figure 2 shows FMR spectra observed at 10.0 K for a $Ga_{1-x}Mn_xAs$ ($x = 0.02$) sample in the four basic applied field orientations: $\mathbf{H}\|[001]$, $\mathbf{H}\|[110]$, $\mathbf{H}\|[1\bar{1}0]$, and **H**||[100]. Note that FMR peaks are observed in all configurations (and persist up to T$_C$). Remarkably, the FMR spectrum and the resonance field (marked by an arrow) depend strongly on orientation of the applied field relative to the crystal directions, indicating that magnetic crystalline anisotropy plays a major role in determining the fields at which the resonances occur.

In order to determine magnetic parameters from the FMR data we will use the Stoner-Wohlfarth model,[26] where the ferromagnetic layer is assumed to consist of a single homogeneous magnetic domain. For a zinc-blende crystal film (such as Ga$_{1-x}$Mn$_x$As) under tetragonal distortion, this leads us to the following expression for the free energy density $F$ in an applied magnetic field $H$:[18,27]

$$F = \frac{1}{2} M \left\{ \begin{array}{l} -2H\left[\cos\theta\cos\theta_H + \sin\theta\sin\theta_H \cos\left(\varphi - \varphi_H\right)\right] + 4\pi M\cos^2\theta - H_{2\perp}\cos^2\theta \\ -\frac{1}{2}H_{4\perp}\cos^4\theta - \frac{1}{2}H_{4\parallel}\frac{1}{4}(3+\cos 4\varphi)\sin^4\theta - H_{2\parallel}\sin^2\theta\sin^2\left(\varphi - \frac{\pi}{4}\right) \end{array} \right\}. \qquad (1)$$

Here the first term describes the Zeeman energy; the second term is the demagnetizing energy (shape anisotropy); $H_{2\perp}$ and $H_{4\perp}$ represent the perpendicular uniaxial and cubic anisotropy fields, respectively; $H_{2\parallel}$ and $H_{4\parallel}$ are the in-plane uniaxial and cubic anisotropy fields, respectively; and the angles $\theta$ and $\varphi$ are defined in Fig. 1.

The resonance condition for any given field orientation can then be obtained from Eq. (1) in the standard manner,[25]

$$\left(\frac{\omega}{\gamma}\right)^2 = \frac{1}{M^2\sin^2\theta}\left[\frac{\partial^2 F}{\partial\theta^2}\frac{\partial^2 F}{\partial\varphi^2} - \left(\frac{\partial^2 F}{\partial\theta\,\partial\varphi}\right)^2\right]. \qquad (2)$$

The equilibrium angle $(\theta, \varphi)$ of the magnetization at the resonance condition is obtained by minimizing the free energy ($\frac{\partial F}{\partial \theta} = 0$ and $\frac{\partial F}{\partial \varphi} = 0$) at a given field orientation $(\theta_H, \varphi_H)$. We do this for the three configurations described in the previous section. Using the coordinates and configurations defined in Fig. 1, for $\varphi = \varphi_H = 45^\circ$ [**H** and **M** in the $(1\bar{1}0)$ plane, referred to as Setup 1], one finds

$$
\left(\frac{\omega}{\gamma}\right)^2 = \{H_R \cos(\theta_H - \theta) + (-4\pi M + H_{2\perp} + H_{4\perp}/2 - H_{4\parallel}/4)\cos 2\theta + (H_{4\perp}/2 + H_{4\parallel}/4)\cos 4\theta\}
$$
$$
\times \{H_R \cos(\theta_H - \theta) + (-4\pi M + H_{2\perp} + H_{4\parallel}/2)\cos^2 \theta + (H_{4\perp} + H_{4\parallel}/2)\cos^4 \theta - H_{4\parallel} - H_{2\perp}\};
$$

$$(3a)$$

for $\varphi_H = 0^\circ$ [**H** and **M** in the $(010)$ plane, referred to as Setup 2],

$$
\left(\frac{\omega}{\gamma}\right)^2 = \{H_R \cos(\theta_H - \theta) + (-4\pi M + H_{2\perp} - 2H_{4\parallel} - H_{2\parallel}/2)\cos^2 \theta + (H_{4\perp} + H_{4\parallel})\cos^4 \theta + H_{4\parallel}\}
$$
$$
\times \{H_R \cos(\theta_H - \theta) + (-4\pi M + H_{2\perp} + H_{4\perp}/2 - H_{4\parallel}/2 - H_{2\parallel}/2)\cos 2\theta + (H_{4\perp}/2 + H_{4\parallel}/2)\cos 4\theta\}
$$
$$
- (H_{2\parallel}/2)^2 \cos^2 \theta;
$$

$$(3b)$$

and for $\theta = \theta_H = 90^\circ$ (**M** and **H** in the $(001)$ plane, i.e., parallel to the film plane, referred to as Setup 3) one obtains,

$$
\left(\frac{\omega}{\gamma}\right)^2 = \{H_R \cos(\varphi - \varphi_H) + H_{4\parallel}\cos 4\varphi - H_{2\parallel}\cos(2\varphi - \pi/2)\}
$$
$$
\times \{H_R \cos(\varphi - \varphi_H) + 4\pi M - H_{2\perp} + H_{4\parallel}(3 + \cos 4\varphi)/4 + H_{2\parallel}\sin^2(\varphi - \pi/4)\}.
$$

$$(3c)$$

Here $\omega$ is the angular frequency of the microwave field and $\gamma = g\mu_B \hbar^{-1}$ is the gyromagnetic ratio, $g$ being the spectroscopic splitting factor, and $\hbar$ the Planck constant. Through these equations the FMR experiments provide a direct measure of magnetic anisotropy fields and of the effective $g$-factor of the Mn-ion/hole complex in the III-Mn-V system. Note that when $\varphi_H = 0^\circ$ (Setup 2), the equilibrium angle $\varphi$ of the magnetization at resonance is not always equal to $\varphi_H$ (0°) due to the finite in-plane anisotropy field $H_{2\parallel}$. However, to simplify the analysis in deriving Eq. (3b), we have ignored the possible small difference between $\varphi$ and $\varphi_H$. Since the resonance field $H_R \gg H_{2\parallel}$, this assumption is reasonable – and is confirmed *a posteriori* by our analysis. Note finally that the terms $4\pi M - H_{2\perp}$ always occur together. For that reason it is customary in calculations to lump $4\pi M - H_{2\perp}$ into a single term, which we will define as $4\pi M_{eff}$.

We have mapped out the FMR fields for our $Ga_{1-x}Mn_xAs$ ($x = 0.02$) sample as a function of magnetic field orientation relative to the crystal axes. The observed resonance fields are shown as data points in Fig. 3 for four temperatures. In each panel (i.e., for each temperature) there are three windows (1, 2, 3) displaying the resonance field $H_R$ as a function of applied field orientation, each window corresponding to one of the three configurations ("setups") described in the section II. Strikingly, as shown in window (1) of each panel, at lower temperatures (T = 10 K) the angular dependence of $H_R$ for Setup 1 is dominated by a four-fold symmetry, but gradually transforms to two-fold symmetry as temperature increases (see T = 30 K). Similar symmetry characteristics are also present (though less obvious) in window (2) of the successive panels in Fig. 3, corresponding to the second out-of-plane configuration (Setup 2). We attribute the observed symmetry behavior to competition between the cubic ($H_{4\perp}$) and the uniaxial ($H_{2\perp}$) anisotropies. From this behavior we infer that at low temperatures the angular dependence of $H_R$ is dominated by the cubic anisotropy field ($H_{4\perp} > H_{2\perp}$); but at higher temperatures the role of uniaxial anisotropy field becomes increasingly important, eventually becoming dominant, as seen in the 30K panel of Fig. 3 ($H_{2\perp} > H_{4\perp}$). At same time one should note that the lowest resonance field at each temperature is always observed when **H** lies in the film plane and parallel to the [100] direction ($\theta_H = 90^o$, $\varphi_H = 0^o$), as was already seen in the spectra in Fig. 2. As a result, the easy axis of this $Ga_{1-x}Mn_xAs$ ($x = 0.02$) sample is always along the [100] or the [010] direction for all temperatures below $T_C$. Moreover, window (3) of each panel in Fig. 3, which shows the angular dependence of $H_R$ for **H** in the (001) plane, directly reveals the existence of the cubic magnetic anisotropy $H_{4\parallel}$ and the uniaxial magnetic anisotropy $H_{2\parallel}$.

By analyzing these in-plane data shown in window (3) via Eq. (3c) we obtain approximate values of $4\pi M_{eff}$, $H_{4\parallel}$ and $H_{2\parallel}$ by first assuming $g = 2.00$ and $H_{4\perp} = 0$. We then use the results obtained under these assumptions as starting parameters to carry out a weighted nonlinear least squares fit [using Eq. (3a) and (3b)] to the FMR positions for the data in the remaining windows (1) and (2), allowing the three parameters ($g$, $4\pi M_{eff}$, and $H_{4\perp}$) to vary. Using parameter so obtained as input for the next iteration, we return to the in-plane data to obtain $4\pi M_{eff}$, $H_{4\parallel}$ and $H_{2\parallel}$ by using the new value of $g$. We iterate these two steps until optimal fitting is achieved and all five parameters ($4\pi M_{eff}$, $H_{4\perp}$, $H_{4\parallel}$, $H_{2\parallel}$, and $g$) converge with subsequent iterations. One can see the excellent fits which have been obtained in this way for all data points, as shown by the solid curves in Fig. 3.

The results for our sample are listed in Table I. Note that the cubic anisotropy field $H_{4\perp}$ is quite large and cannot be neglected. We attribute this prominence of $H_{4\perp}$ to the small value of the compressive strain existing in the sample due to the small Mn concentration ($x = 0.02$). One should note for completeness that some anisotropy of the $g$-factor is also to be expected in this type of system. However, a fit obtained using an anisotropic *g*-factor cannot be distinguished from the fit with an isotropic value of *g*. We have therefore accepted an isotropic *g*-factor as an adequate approximation. As listed in Table I, the effective *g*-factors obtained in this analysis are smaller than the value of 2.00 characteristic of isolated $Mn^{++}$ ions. This can only be attributed to a contribution from the magnetic moment of the holes to the collective FMR precession.[24] In summary, FMR data and its analysis presented in this section unambiguously establish the magnetic anisotropy fields for this $Ga_{1-x}Mn_xAs$ (*x* = 0.02) film, as well as a reduction of the *g*-factor that reflects the formation of the coupled complex of the local Mn ions and the free holes that jointly represent the total magnetic moment of the III-Mn-V system as a whole.

### 3.2. Process of perpendicular magnetization reversal

The reversal of magnetization has been thoroughly explored both experimentally and theoretically in thin ferromagnetic metal layers (e.g., iron) for the case when the external field is applied *in the layer plane*.[28, 29] Here we focus on the situation where an external magnetic field is applied perpendicular to the sample surface ($\theta_H = 0º$), and again use the single domain model as in Refs. [12] and [29]. Since $H_{4\parallel} > 0$ (see Table I), [100] and [010] are the easy axes of magnetization, and we can then take $\varphi = \varphi_H = 0º$ or 90º, i.e., the magnetization will be confined to either (010) or (100) planes, which we have referred to as the easy planes of magnetization. In this case the free-energy density *F* simplifies to

$$F = \frac{1}{2}M[-2H\cos\theta + (4\pi M_{eff} + \frac{1}{2}H_{2\parallel})\cos^2\theta - \frac{1}{2}H_{4\perp}\cos^4\theta - \frac{1}{2}H_{4\parallel}\sin^4\theta]. \qquad (4)$$

In terms of the normalized perpendicular magnetization $\frac{M_Z}{M} = \xi \equiv \cos\theta$, Eq. (4) can be expressed as a quartic expression

$$F = -\frac{1}{2}M[2H\xi - (4\pi M_{eff} + \frac{1}{2}H_{2\parallel} + H_{4\parallel})\xi^2 + (\frac{1}{2}H_{4\perp} + \frac{1}{2}H_{4\parallel})\xi^4 + \frac{1}{2}H_{4\parallel}]. \qquad (5)$$

By minimizing the free energy $F$ with respect to $\theta$, the equilibrium angle of the magnetization can be obtained by solving the equation

$$\frac{dF}{d\theta} = M\sin(\theta)[H - (4\pi M_{eff} + \frac{1}{2}H_{2\parallel} + H_{4\parallel})\xi + (H_{4\perp} + H_{4\parallel})\xi^3] = 0, \qquad (6)$$

with the additional condition

$$\frac{d^2F}{d\theta^2} = M\cos(\theta)[H - (4\pi M_{eff} + \frac{1}{2}H_{2\parallel} + H_{4\parallel})\xi + (H_{4\perp} + H_{4\parallel})\xi^3]$$
$$+ M\sin^2(\theta)[(4\pi M_{eff} + \frac{1}{2}H_{2\parallel} + H_{4\parallel}) - 3(H_{4\perp} + H_{4\parallel})\xi^2] > 0, \qquad (7)$$

to guarantee that the solution is a local minimum of the energy.

If only a single homogeneous magnetic domain is present and thermal fluctuations are not considered (i.e., T = 0 K), the solutions of Eqs. (6) and (7) are sufficient for describing the magnetization reversal process and the accompanying hysteresis loop. To illustrate this, in Fig. 4 we plot the free energy per Mn$^{++}$ ion as a function of the angle of magnetization $\theta$ for several values of the applied magnetic field calculated using the anisotropy fields for $T$ = 10 K listed in Table I. The free energies are calculated for applied fields $H$ = -400, -43, 400, 800, 1200, 1442, and 1600 Oe, and $5\mu_B$ per Mn$^{++}$ ion is assumed. As seen in Fig. 4, as a consequence of competition between cubic and uniaxial anisotropy terms, for most values of the applied field $H$ there are two local energy minima separated by an energy barrier. The lowest minimum occurs at the special condition of $\theta = 0^\circ$, i.e., when the magnetization $\boldsymbol{M}$ is parallel to the applied field $\boldsymbol{H}$. This $\theta = 0^\circ$ minimum, which requires a high field, becomes more shallow when the field decreases, and eventually completely disappears when the applied field is reduced below a certain value $H_{N1}$ given by

$$H_{N1} = 4\pi M_{eff} + \frac{1}{2} H_{2\parallel} - H_{4\perp}. \qquad (8)$$

For parameters used in the present calculation $H_{N1} = -43$ Oe (see solid curve marked $H_{N1}$ in Fig. 4).

The second energy minimum corresponds to one of three solutions of the cubic equation

$$H = (4\pi M_{eff} + \frac{1}{2} H_{2\parallel} + H_{4\parallel})\xi - (H_{4\perp} + H_{4\parallel})\xi^3, \qquad (9)$$

which additionally needs to satisfy the condition $4\pi M_{eff} + \frac{1}{2} H_{2\parallel} + H_{4\parallel} > 3(H_{4\perp} + H_{4\parallel})\xi^2$. This solution shows that $M_Z$ *monotonically* increases (decreases) when the applied field increases (decreases). This minimum occurs at $\theta = 90º$ for $H = 0$ (i.e., *M* is along the easy axis of magnetization). As the magnetic field increases, the angle $\theta$ corresponding to this second minimum gradually decreases (see inset on lower right in Fig. 4), i.e., the entire magnetic moment rotates coherently and continuously toward the field direction $\theta_H = 0º$. One can show from Eq. (9) that at low field $M_Z$ increases *monotonically* with increasing field at the rate

$$\frac{dM_Z}{dH} = \frac{M}{4\pi M_{eff} + \frac{1}{2} H_{2\parallel} + H_{4\parallel}}, \qquad (10)$$

but as the field continues to grow, the second energy minimum will eventually disappear when $H$ exceeds the value $H_{N2}$ given by

$$H_{N2} = \frac{2\sqrt{3}}{9}(H_{4\perp} + H_{4\parallel})(\frac{4\pi M_{eff} + \frac{1}{2}H_{2\parallel} + H_{4\parallel}}{H_{4\perp} + H_{4\parallel}})^{\frac{3}{2}}. \qquad (11)$$

It is easy to see from Fig. 4 that when the applied field approaches $H_{N2}$, the orientation of the magnetic moment will undergo a rapid change, and eventually will "snap" to coincide with the applied field direction at $H = H_{N2}$ (see solid curve marked $H_{N2} = 1442$ Oe in the figure). At that point the free energy has only one minimum, that for $\theta = 0º$.

As we begin the field reversal process, i.e., as the applied field is decreased and approaches $H_{N1}$ from above, the local minimum at $\theta = 0º$ will become increasingly shallow, as already described, and will eventually disappear. The magnetic moment then shifts from the field direction ($\theta = 0º$) to another orientation corresponding to another free energy minimum. The shift of the local free energy minimum described above thus generates a hysteresis loop for the magnetization $M_Z$, which is plotted as a solid curve in the upper inset in Fig. 4. However, the hysteresis loop generated in this way does not agree with experimental results, indicating that other mechanisms must be considered to establish the correct model for the observed behavior.

Note that, since at a finite temperature thermal fluctuations can excite the system over energy barriers to a lower energy minimum, the actual area of the hysteresis loop will be reduced dramatically or even disappear.[30] For example, as plotted in Fig. 4, at $H = H_T = 1100$ Oe, the two energy minima (at $\theta = 0º$ and $\theta = 71º$) are equal. They are separated by an energy barrier of 0.003 mev, which is far less than the magnitude of the thermal fluctuation energy $k_BT$ even at very low temperature (e.g., $k_BT = 0.86$ meV at 10 K). Since the exchange integral $J$ of $Ga_{1-x}Mn_xAs$ is also small ($\sim 0.1$ meV), the "snap" of the spin orientation actually occurs around $H_T$ (where the two energy minima are equal) via tunneling, even through a finite energy barrier between the two minima is still present. In this situation, $H_{N1}$ and $H_{N2}$ are actually not the observed hysteresis "snap" fields. Note that in actuality the tunneling through the energy barrier takes place via non-coherent spin switching, similar to that discussed for the in-plane magnetization reversal.[12]

Although an analytical solution for $H_T$ can be found by solving Eqs. (5) – (7), it is very complicated and not straightforward. Here we give a more convenient simple approximation by ignoring third order terms of $\xi$:

$$H_T \approx (4\pi M_{eff} + \frac{1}{2}H_{2\parallel} + H_{4\parallel}) - \sqrt{\frac{1}{2}(4\pi M_{eff} + \frac{1}{2}H_{2\parallel} + H_{4\parallel})(H_{4\perp} + H_{4\parallel})}. \qquad (12)$$

Using the same approximation, one can show that the height of energy barrier separating the two local energy minima is about

$$\Delta F \approx \frac{1}{2}M[(2-\sqrt{2})H_T - \frac{1}{2}(4\pi M_{eff} + \frac{1}{2}H_{2\parallel} + H_{4\parallel}) + \frac{3}{4}(\frac{1}{2}H_{4\perp} + \frac{1}{2}H_{4\parallel})]. \qquad (13)$$

The above results can be applied to find analytic expressions in a ferromagnetic film with a known magnetic anisotropy. Before we do this (in the next section), in Fig. 5 we will first summarize the above discussion by calculating representative magnetization curves as a function of applied field for several physically important limits.[31] In Fig. 5(a), under the condition $4\pi M_{eff} + \frac{1}{2}H_{2\parallel} > 3H_{4\perp} + 2H_{4\parallel}$, the solutions for $H_{N2}$ and $H_T$ do not exist and the field $H_{N1}$ simply represents the field when saturation of the magnetization is attained. In this case there is no hysteresis, and no jumps in magnetization take place; i.e., the spins rotate *coherently as a single unit*, without non-coherent spin switching. If furthermore $H_{4\parallel}$ and $H_{4\perp}$ are *negligibly* small, the magnetization curve will be a straight line before the field reaches $H_{N1}$. This form of magnetization (associated with the inequality given above) is usually observed in $Ga_{1-x}Mn_xAs/GaAs$ systems with large Mn concentration ($x > 0.05$), since the built-in compressive strain is in this case quite large, and properties determined by the out-of-plane cubic symmetry terms of $Ga_{1-x}Mn_xAs$ are completely overshadowed by those arising from the uniaxial distortion of the material induced by the lattice mismatch.

Figure 5(b) corresponds to a typical $Ga_{1-x}Mn_xAs/GaAs$ system with _low_ Mn concentration ($x < 0.03$). In this case (which can be described by the condition $\frac{1}{2}(H_{4\perp} - H_{4\parallel}) < 4\pi M_{eff} + \frac{1}{2}H_{2\parallel} < 3H_{4\perp} + 2H_{4\parallel}$), uniaxial anisotropy terms induced by the weaker compressive strain (smaller lattice mismatch) are of the same order as the cubic anisotropy terms. As shown in that panel by the solid curve, the magnetization reversal then consists of two steps: a coherent spin rotation for $-H_T < H < H_T$, and a non-coherent spin switching through the energy barrier at $H = \pm H_T$, which is due to thermal fluctuations.

Finally, Fig. 5(c) illustrates the case when $4\pi M_{eff} + \frac{1}{2}H_{2\parallel} < \frac{1}{2}(H_{4\perp} - H_{4\parallel})$. In this case the [001] direction (the normal to layer plane) will be the easy axis of magnetization. The magnetization will then switch reversibly around zero field, as shown by the solid curve in Fig. 5(c). This is typical behavior for the $Ga_{1-x}Mn_xAs/Ga_{1-y}In_yAs$ systems, where for a sufficiently high In concentration $y$ the tensile strain is present in the $Ga_{1-x}Mn_xAs$ layer, resulting in the inequality $4\pi M_{eff} \equiv 4\pi M - H_{2\perp} < 0$.

It should be mentioned that in reality, as the applied field approaches $H_T$, some portions of the total magnetic moment may tunnel ahead of all others through the energy barrier as a result of thermal fluctuations. The specimen is therefore comprised *at that* *instant* of small regions (domains) with different magnetic moment orientations. With this domain effect present the magnetization reversal will not be *reversible* at $H = \pm H_T$, leading to *double hysteresis loops*, especially as T → 0K. It is easy to see that the width of the two hysteresis loops will increase as T → 0K, moving toward the limiting fields $H_{N1}$ and $H_{N2}$ around $H_T$, as shown by the dashed curves in Fig. 5(b). Indeed, such complex hysteresis behavior has been observed in both polar magneto-optic Kerr effect (MOKE)[32] and in Hall measurements. This aspect of magnetization reversal will be discussed in terms of domain nucleation and expansion later in the paper.

### 3.3. Multi-step spin switching

It is well known that the anomalous Hall effect (AHE) dominates Hall measurements in ferromagnetic the $Ga_{1-x}Mn_xAs$ system, particularly at low temperatures.[8] Since AHE, $\rho_{Hall}$, is proportional to the perpendicular component of the magnetization $M_Z$, in this paper we make use of $\rho_{Hall}$ measurements to study $M_Z$ vs. $H$. In Fig. 6 we show Hall data observed at several temperatures for one specific $Ga_{1-x}Mn_xAs$ ($x = 0.02$) film, cleaved from the same layer as that used in the FMR study discussed above. The sample configuration is shown in Fig. 7(a). The longer dimension of the Hall bar sample, and hence the direction of the current **I**, was chosen along the [1$\bar{1}$0] crystallographic direction. Measurements were performed with the magnetic field **H** normal to the (001) plane [i.e., along $z$ in Fig. 7(a)]. Note that the temperatures were chosen to be the same as those used in the FMR, discussed in the previous section. The experimental data are plotted as open-square symbols in Fig. 6. The solid curves showing the jump at field $H_T$ are calculated using the analytical expressions derived in the previous section, along with the anisotropy parameters listed in Table I. Although there is a small but unmistakable hysteresis loop around the field $H_T$, the data shown in the figure unambiguously confirm that the saturation of $M_Z$ occurs at $H = \pm H_T$ (marked as 3 and 8). Clearly the magnetic moment does not "wait" for the free energy barrier to disappear, but tunnels through the free energy barrier as soon as the two local energy minima become equal (see Fig. 4).

It should be mentioned that in reality the magnetic field is usually not perfectly aligned with the [001] direction. Thus there always exists a magnetic field "shadow" in the layer plane, and consequently a slight difference between the $xz$ and the $yz$ plane. As a result, the Hall resistivity $\rho_{Hall}$ actually contains two contributions: one from AHE, which is proportional to the perpendicular magnetization $M_Z$; and one from the anisotropic magneto-resistance (AMR)[33] in the layer plane, referred to as the planar Hall effect (PHE). Thus $\rho_{Hall}$ in a $Ga_{1-x}Mn_xAs$ film depends strongly on the relative orientation of the *in-plane* magnetization with respect to the direction of the current.[3,20] As shown in Fig. 6, at T = 10 and 15K three jumps (and even four at T = 20K) are observed in the Hall resistivity $\rho_{Hall}$ when the applied field $H$ sweeps up from -2000 Oe to 2000 Oe, or *vice versa*. For example, two jumps marked 3 and 9 are attributed to the AHE variation caused by the spin switching at $H = \pm H_T$ when $H$ sweeps up from -2000 Oe to 2000 Oe; and 4 and 8 are the corresponding jumps when $H$ is swept *down* from 2000 to -2000 Oe. The remaining jumps, marked 1 and 6 in Fig. 6, are attributed to the PHE contribution, and correspond to spin switching between the two easy planes (100) and (010) of magnetization.

One should note that all jumps seen in Fig. 6 are related to domain nucleation and spin switching through an intermediate state. To see this, let us concentrate on the curve for T = 10 K in Fig. 6 and the corresponding schematic plot in Fig. 7. Note that in Fig. 7(a) there are *two* easy planes of magnetization: the *xz* plane [i.e., the (010) plane] and the *yz* plane [i.e., the (100) plane]. When the applied field is zero, we assume that magnetic moment **M** is along one of the easy axes in the layer plane (e.g., the [$\bar{1}00$] direction, i.e., along the –*x* direction in the *xz*-plane). When the field begins to increase, **M** tilts toward the field direction, initially with a slope $\frac{dM_Z}{dH} = M/(4\pi M_{eff} + \frac{1}{2}H_{2\parallel} + H_{4\parallel})$ in the *xz* plane (marked 0 in Figs. 6 and 7b). As the field continues to increase, the first jump (marked 1 in Figs. 6 and 7) observed at a small positive value (~500 Oe) is caused by the presence of a small in-plane projection ("shadow") of the applied field that leads to a preference of one easy plane over the other. The jump thus indicates a 90º spin switching from the *xz* plane to the *yz* plane (the other easy plane), and originates from PHE. With further increase in the field, the direction of **M** continues to tilt further toward the direction of **H**, while remaining in the *yz* plane (marked 2), eventually reaching the second jump at $H_T$ (~1100 Oe), that corresponds to a 71º switching of the magnetization (by tunneling through the barrier) from near the layer plane to the field direction, marked 3 in Figs. 6 and 7. As the field continues to increase beyond that value, **M** remains aligned with applied field.

During the down-sweep, when the field decreases to 1100 Oe from above the magnetization **M** jumps back to the *xz* plane with another 71º spin switching, marked 4 in Figs. 6 and 7. When the applied field is swept from 1000 Oe to -500 Oe, **M** is rotated from the direction above the layer plane to below the layer plane in the *xz* plane (marked 5 in Figs. 6 and 7). When the field reaches the value of -500 Oe, the PHE jump marked 6 in Figs. 6 and 7 is observed, caused by a 90$^o$ domain rotation from the $xz$ plane to the $yz$ plane. As field continues to -2000 Oe, we observe the magnetic moment rotation (7) and switching (8) which are the reverse-field analogues of the processes marked 2 and 3, already discussed. When the field returns from -2000 Oe to zero, one more jump (marked as 9, reverse-field analogue of 4) is observed.

Note that this complicated magnetization reversal process consists of both coherent spin rotation (0, 2, 5, and 7) and non-coherent spin switching (1, 3, 4, 6, 8, and 9). In particular, all non-coherent spin switchings create the clear abrupt changes in the Hall resistivity $\rho_{\text{Hall}}$ (as well as similar abrupt changes with sheet resistivity, not shown here), which can be exploited in the design of potential devices involving magnetic logic operations.

Importantly, by combining these Hall resistivity data with the data from the *standard* PHE measurements described in Ref. [3], one can obtain all four magnetic anisotropy fields using only magneto-transport measurements. In this procedure, the data from PHE measurements will first yield the values of the two in-plane magnetic anisotropy fields $H_{2\parallel}$ and $H_{4\parallel}$. After that the slope of the magnetization rotation at low field (i.e., processes 0, 2, 5, and 7) and the spin switching field $H_T$ [i.e., the average field at which processes 3 and 4 (or 8 and 9) take place] can be used to obtain the two remaining magnetic anisotropy fields $H_{2\perp}$ and $H_{4\perp}$ of the specimen under study.

The anisotropy fields at various temperatures obtained from magneto-transport (below 10 K) and FMR (above 10 K) of the $Ga_{1-x}Mn_xAs$ ($x = 0.02$) film are plotted in Fig. 8. For completeness, we also plot the normalized saturation magnetization deduced from the magneto-transport data by using the Arrott plot and by assuming that AHE is dominated by skew scattering – i.e., that M $\propto$ $R_{Hall}$/$R_{sheet}$, where $R_{Hall}$ and $R_{sheet}$ are the Hall and sheet resistances, respectively, when **H** is applied perpendicular to the layer plane. As shown in Fig. 8, the magnetic anisotropy fields are strongly temperature-dependent. It is interesting that the cubic anisotropy fields are not necessarily equal to each other; however they all decrease quickly with increasing temperature. Note that the perpendicular uniaxial anisotropy field (which is contained in $4\pi M_{eff}$) is not decreasing monotonically as the temperature is increasing, illustrating the complex nature of magnetic anisotropy in the $Ga_{1-x}Mn_xAs$ system, which depends strongly on the hole concentration, magnetization, and temperature.[34]

### 3.4. Domain nucleation and expansion

One can see from Fig. 6 that the fields where the first jumps occur in $\rho_{Hall}$ (marked 1 and 6) increase rapidly as the temperature decreases, while the jump corresponding to $H_T$ remains fixed. Projecting this trend, one may expect that at very low temperatures the value of the field corresponding to steps 1 and 6 will eventually exceed $H_T$. As a result, the entire process of perpendicular magnetization reversal is confined to one easy plane (e.g., $xz$ plane) exactly as observed at T = 1.45 K, shown in Fig. 9. Note the appearance of two obvious hysteresis loops centered at $H_T$ and $-H_T$, each with a coercive field $H_{C1}$ and $H_{C2}$. It is perhaps unexpected for Hall data observed on a single magnetic film to exhibit such double hysteresis loops positioned symmetrically around zero field. Similar magnetization profiles were observed earlier in polar MOKE measurements.[32] Given the magnetic anisotropy data in Fig. 8, we obtain analytic expressions for the magnetization curves at T = 1.45 K. The calculated results are plotted in Fig. 9 as solid curves. Clearly, apart from the absence of hysteresis loops, the calculated solid curves reproduce the experimental result very well.

In the inset in Fig. 9 we have plotted the free energy per Mn ion (assuming $5\mu_B$ per Mn ion) as a function of the angle of magnetization $\theta$ at three fields $H_T$, $H_{C1}$ and $H_{C2}$. Clearly all the plots show the presence of an energy barrier between the two local minima for all curves. Note that at $H = H_{C1}$ or $H_{C2}$ the free energy corresponding to the final state of magnetization is lower than that of the initial state by an increment $\Delta E$. We attribute this energy increment $\Delta E$ to the formation energy of the domain wall. Due to the shallow energy barrier and the small value of the exchange energy $J$, some spins in the system can flip to the lower state via thermal fluctuations. These flipped spins can be regarded as domain nuclei, and we therefore refer to this process as domain nucleation. Since a domain wall needs to be created around these domain nuclei, this process of domain wall formation will require additional energy. As soon as domain nucleation occurs in the system (including the formation of domain walls), thermal fluctuations together with the applied magnetic field will force the domain "nuclei" with lower energy to expand rapidly throughout the entire specimen. Thus the magnetic moment of the specimen as a whole will eventually approach a final state that corresponds to the lowest energy, and the spins comprising the system will then again behave coherently. The phenomenon of domain nucleation and expansion has already been observed in the in-plane magnetization reversal process by using magneto-optical domain imaging.[10] However, to observe similar effect in the reversal process of perpendicular magnetization, one should use imaging experiments with a higher resolution (~0.1 μm), due to the much smaller domain sizes that characterize the perpendicular magnetization reversal process.[35]

Finally, we use Kittel's Bloch wall model[36] to calculate the energy per unit area of the domain wall, i.e., we estimate the sum of the contributions from exchange and anisotropy energies associated with the wall formation, $\sigma_{\text{w}} = \sigma_{\text{ex}} + \sigma_{\text{anis}}$. Note that a detailed theory of Bloch domain walls in III$_{1-x}$Mn$_x$V systems has already been discussed earlier by Dietl *et al.*[37] Here we use a rather simplified intuitive picture, which retains the essential points of Ref. [37] without losing physical insight. Specifically, the picture assumes that the total spin orientation within the domain wall changes by the angle $\pi/2$, and that the thickness of the domain wall consists of $N$ unit cells of Mn ion sub-lattice. The exchange energy $\sigma_{\text{ex}}$ is then given by $\sigma_{\text{ex}} \approx \pi^2 J S^2 Q / 4 N d^2$. Here $S$ is the spin quantum number of the Mn ion; $Q$ is an integer with values 1, 2, or 4 for sc, bcc, and fcc lattices, respectively; and $d$ is the average lattice constant of the Mn ion sub-lattice, $d = a/x^{1/3}$, where $a$ is the lattice constant of Ga$_{1-x}$Mn$_x$As, and $x$ is the Mn concentration. The anisotropy energy $\sigma_{\text{anis}}$ is of the order of the free energy barrier times the thickness $Nd$, or $\sigma_{\text{anis}} \approx Nd\Delta F$. By using these expressions in $\sigma_{\text{w}}$ defined above and finding the minimum of $\sigma_{\text{w}}$ with respect to $N$, i.e., $\dfrac{\partial \sigma_{\text{w}}}{\partial N} = 0$, we obtain

$$N = \frac{\pi}{2}\sqrt{\frac{JS^2 Q}{\Delta F d^3}}, \qquad (14)$$

and

$$\sigma_{\text{w}} = \pi\sqrt{\frac{\Delta F J S^2 Q}{d}}. \qquad (15)$$

Taking $Q = 4$ for the zinc blende structure, $x = 0.02$, $J = 0.1$ meV,[38] $S = 5/2$, and $\Delta F = 0.0058$ meV per Mn ion from Fig. 9, we obtain $N \sim 16$ (i.e., the domain wall thickness $Nd \approx 34$ nm) and $\sigma_{\text{w}} = 0.028$ erg/cm$^2$. For comparison, these values are much smaller than the corresponding values for iron, $N \approx 300$ and $\sigma_w = 1$ erg/cm$^2$. As an example, assuming the initial domain nucleus to have a square shape, we obtain the size of such a domain nucleus to be approximately $4\sigma_w/\Delta E \approx 0.4$ μm. This estimate for the domain nucleus is much smaller than the in-plane domain sizes observed in the $Ga_{1-x}Mn_xAs/GaAs$ system by direct magneto-optical imaging (several hundred μm),[10] but is only a little less than the domain size observed in the $Ga_{1-x}Mn_xAs/Ga_{1-y}In_yAs$ combination, where the magnetization is normal to the layer plane.[35] In this latter case a stripe-shaped domain pattern was observed by a scanning Hall probe microscope, with domain widths of only 1.5-6.4 μm.

## 4. CONCLUDING REMARKS

Understanding of magnetic anisotropy and its manipulation is expected to be an important factor in designing future devices. In this context, the magnetization reversal process plays an especially important role and must therefore be well understood. In this paper, the magnetic anisotropy fields have been mapped out using detailed FMR measurements. We have incorporated these anisotropy fields into a simple model which we use to successfully describe perpendicular magnetization reversal as observed in magnetotrasport experiments involving AHE and PHE. Here we should note that the competition between the cubic and uniaxial anisotropy terms which accompanies the magnetization reversal process is made especially evident in $Ga_{1-x}Mn_xAs$ specimens with low Mn concentration, such as those used in the present study. All evidence clearly points to the key role which magnetic anisotropy plays in this process. Furthermore, our analysis indicates that the multi-step magnetization jumps and the double hysteresis loops accompanying the reversal of perpendicular magnetization originate from irreversible non-coherent spin switching that results from domain nucleation and expansion. An important factor in this process is the formation of domain walls. Here we have used our understanding of magnetic anisotropy obtained from FMR and magnetotransport data, together with Kittel's Bloch wall model, to estimate domain wall thickness and domain size in $Ga_{1-x}Mn_x As$ with low values of $x$. All these processes taken together have given a consistent picture of perpendicular magnetization reversal in $Ga_{1-x}Mn_x As$, indicating the complex series of consecutive steps which must be taken into account in describing this process.

## ACKNOWLEDGMENT

The authors thank R. A. Lukaszew for valuable discussions. This work was supported by the DARPA/SpinS Program through the Office of Naval Research, and by NSF Grant DMR02-10519.

earlier report the authors did not take into account the effect of non-coherent spin switching through the energy barrier.

**TABLE I:**    Magnetic anisotropy fields and *g*-factors derived from the analysis of the angular dependence of FMR fields at various temperatures

| T (K) | $4\pi M_{eff}$ (Oe) | $H_{4\parallel}$ (Oe) | $H_{4\perp}$ (Oe) | $H_{2\parallel}$ (Oe) | $g$ |
|-------|------------------|------------------|------------------|------------------|------------------|
| 10 | 2083±64 | 1985±71 | 1826±78 | -608±80 | 1.87±0.02 |
| 15 | 2174±35 | 1368±62 | 1545±46 | -534±71 | 1.86±0.01 |
| 20 | 2043±52 | 940±71 | 892±66 | -518±69 | 1.89±0.02 |
| 30 | 1555±37 | 295±63 | 312±46 | -314±68 | 1.90±0.01 |

**Figure Captions:**

Fig. 1  Coordinate system and the three experimental configurations used in this paper. The orientation of the dc magnetic field **H** (given by $\theta_H$ and $\varphi_H$) can be varied continuously in the $(1\bar{1}0)$, (010) and (001) planes (Setups 1, 2 and 3, respectively). The corresponding equilibrium orientations of the magnetization **M** are given by $(\theta, \varphi)$.

Fig. 2  FMR spectra observed at T = 10K for $Ga_{0.98}Mn_{0.02}As$ in four high symmetry configurations: **H**||[001], **H**||[110], **H**|| $[1\bar{1}0]$, and **H**||[100]. The arrows point to the resonance position. Note that FMR in the **H**||$[1\bar{1}0]$ configuration is measured in Setup-3 configuration, where the FMR signal is very weak.

Fig. 3  Angular dependence of FMR fields for the $Ga_{0.98}Mn_{0.02}As$ specimen at four different temperatures, shown as four panels. In each panel, window (1) corresponds to the dc magnetic field **H** and magnetization **M** in the $(1\bar{1}0)$ plane, i.e., Setup 1; window (2) to **H** and **M** in the (010) plane, Setup 2; and window (3) to **H** and **M** in the (001) plane, Setup 3. The solid curves in the figure are theoretical fits to the FMR positions $H_R$.

Fig. 4  Free energies (solid and dotted curves) per Mn ion, calculated assuming $5\mu_B$ per Mn ion, plotted as a function of the angle of magnetization $\theta$ at several fields (from -400 Oe to 1600 Oe) applied normal to the layer plane ($\theta_H = 0º$). The initial state of magnetization (at $H = 0$, before the field is applied) is taken to be at $\theta = 90º$. As the applied field increases, the local energy minimum moves continuously to the field

direction $\theta_H = 0°$, as indicated by the arrow on the curves at the right. Note that for the solid curves calculated for $H = H_{N2}$ (1442 Oe) and for $H = H_{N1}$ (0 Oe), the energy barrier between the two local minima disappears, and the magnetization can then slide to the lowest energy minimum on the corresponding curve. The derived perpendicular magnetization hysteresis loop is plotted as an inset. The free energy (solid) curve marked $H_T = 1100$ Oe, also plotted in the figure, corresponds to the field at which two local minima are equal, so that tunneling from one orientation of $\boldsymbol{M}$ to the other can occur.

Fig. 5 Magnetization curves calculated for three parameter regions: (a) $4\pi M_{eff} + \frac{1}{2} H_{2\parallel} = 2000$ Oe, $H_{4\parallel} = H_{4\perp} = 200$ Oe; (b) $4\pi M_{eff} + \frac{1}{2} H_{2\parallel} = 2000$ Oe, $H_{4\parallel} = H_{4\perp} = 1500$ Oe; and (c) $4\pi M_{eff} + \frac{1}{2} H_{2\parallel} = -1000$ Oe, $H_{4\parallel} = 500$ Oe, and $H_{4\perp} = 200$ Oe. In (b) and (c) the dashed curves represent hysteresis loops calculated for $T = 0$ K in the Stone-Wohlfarth limit. The solid curves are expected for experiments carried out at finite temperatures, because thermal fluctuations can then excite the system over energy barriers to lower energy minima. In actual experiments, however, a small hysteresis loop should also be observed around each jump in the magnetization due to the existence of domain wall energy.

Fig. 6 Hall effect data $\rho_{Hall}$ at four different temperatures for the $Ga_{0.98}Mn_{0.02}As$ specimen. Magnetic field $\mathbf{H}$ is applied along the hard axis of magnetization, $\mathbf{H}\|[001]$. The numbers on the bottom curve correspond to specific processes that accompany magnetization reversal, as described in the text.

Fig. 7 (a) Schematic diagram of the sample configuration used for the $\rho_{\mathrm{Hall}}$ measurements. (b) Illustration of magnetic moment rotation and switching for T =10 K for two easy planes of magnetization. The open or solid arrowheds indicate the magnetic moment **M** pointing above or below the layer plane, respectively.

Fig. 8 Magnetic anisotropy fields $H_{2\parallel}$, $H_{4\parallel}$, $H_{2\perp}$ and $H_{4\perp}$ plotted as a function of temperature for the $Ga_{0.98}Mn_{0.02}As$ specimen. The anisotropy fields below 10 K are obtained from magneto-transport, and those above 10 K from FMR. The magnetization data obtained from magneto-transport data via the Arrott plot is also shown in the figure.

Fig. 9 Hall effect data $\rho_{\mathrm{Hall}}$ observed at T =1.45 K on the $Ga_{0.98}Mn_{0.02}As$ specimen. The magnetic field **H** is applied along the hard axis of magnetization, **H**∥[001]. Inset: free energies per Mn ion (assuming $5\mu_B$ per Mn ion) plotted as a function of the angle of magnetization $\theta$ for the three special fields: $H_T$, $H_{C1}$ and $H_{C2}$.

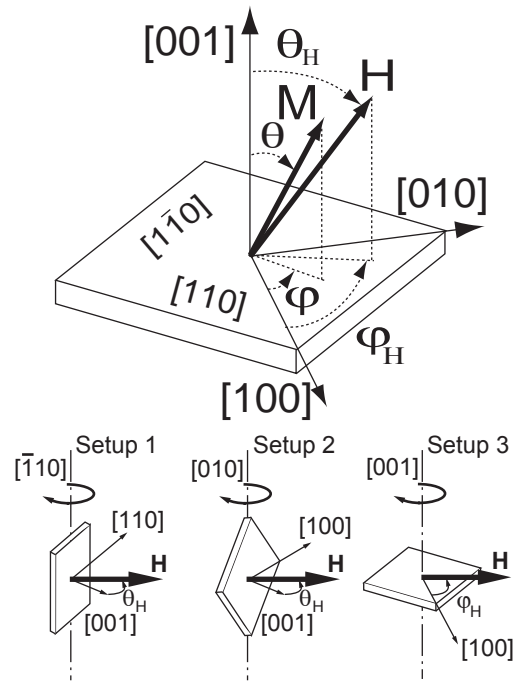

Fig. 1  X. Liu *et al.*

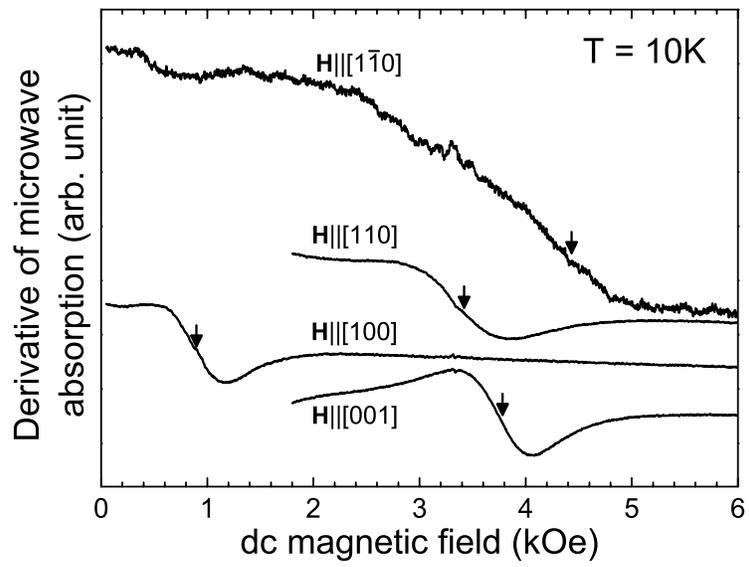

Fig. 2 X. Liu *et al.*

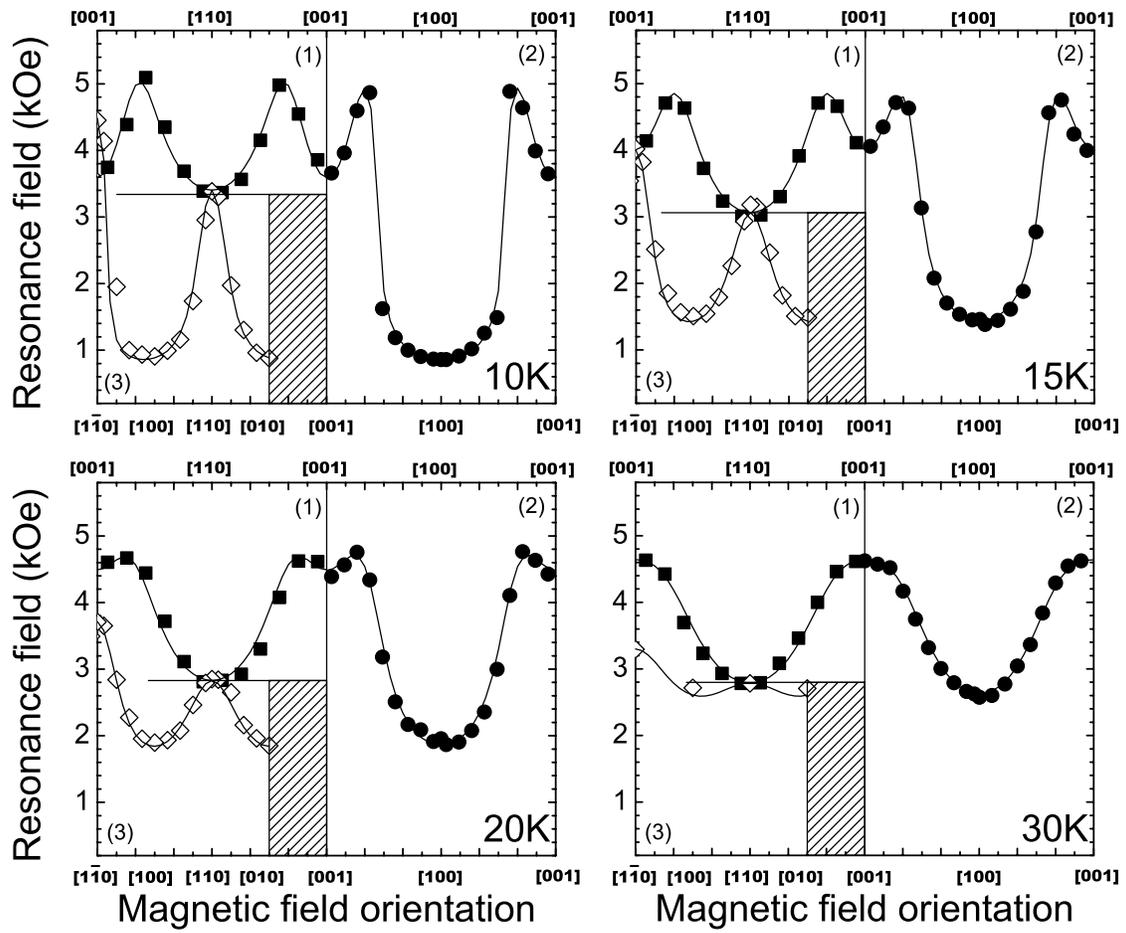

Fig. 3  X. Liu *et al.*

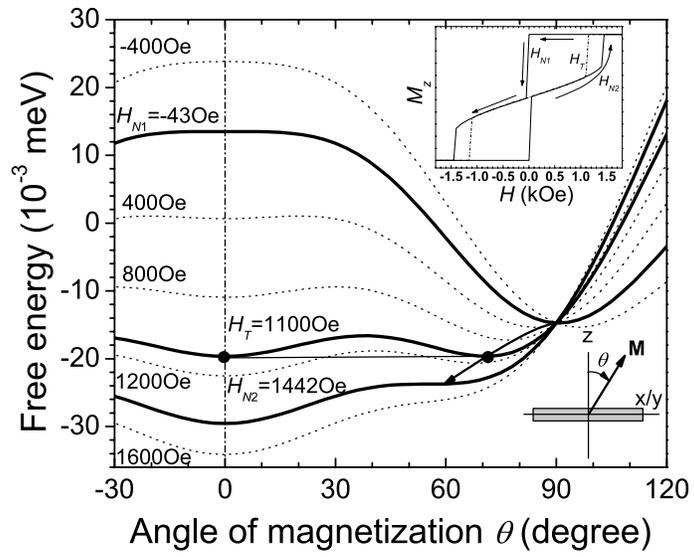

Fig. 4   X. Liu *et al.*

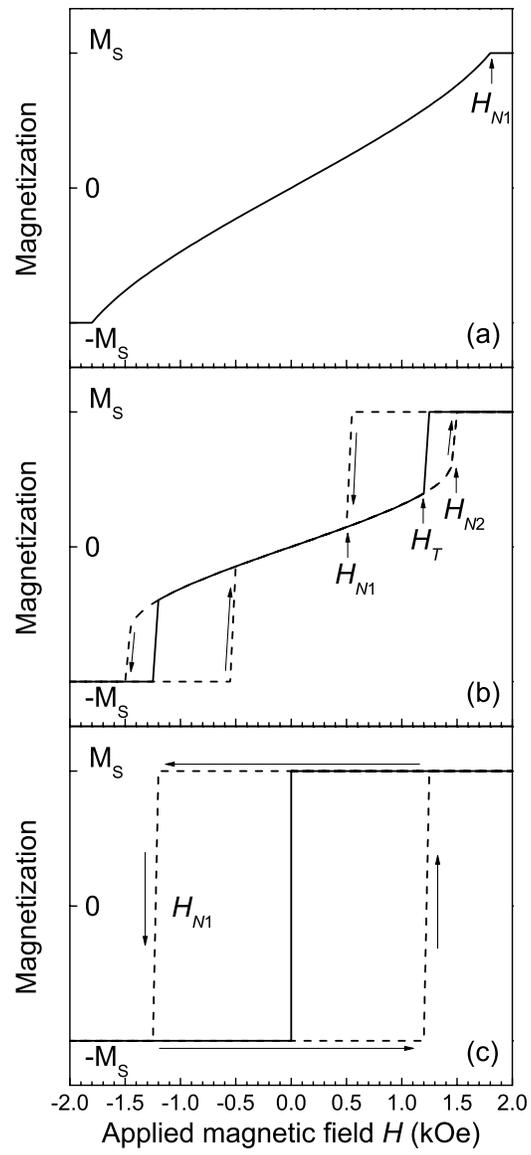

Fig. 5  X. Liu *et al.*

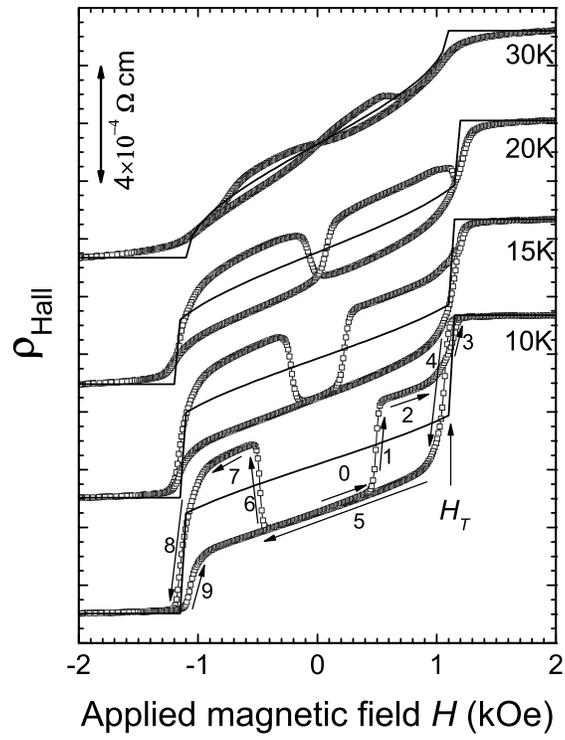

Fig. 6   X. Liu *et al.*

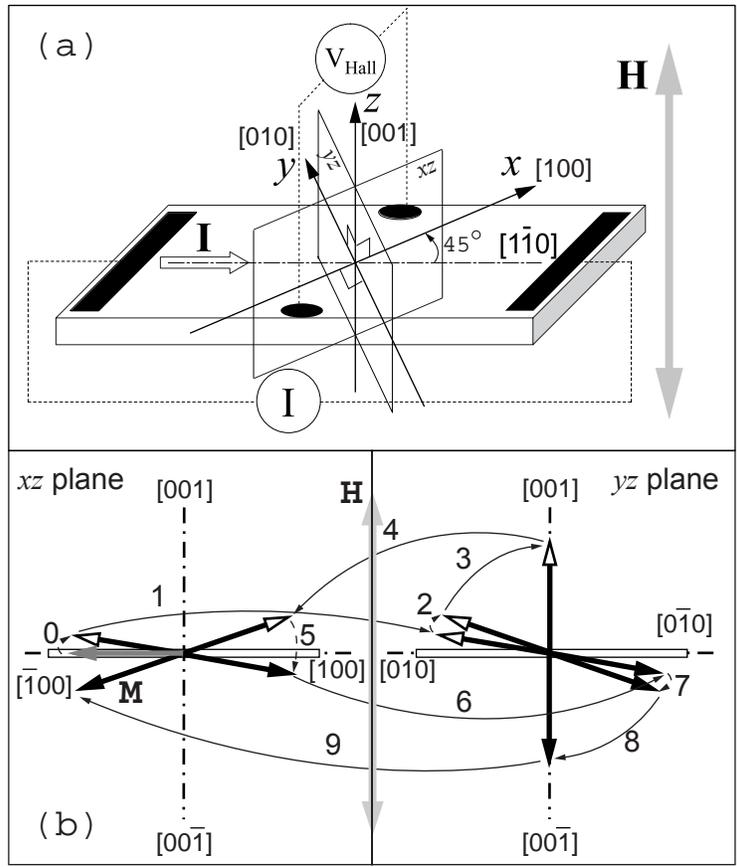

Fig. 7  X. Liu *et al.*

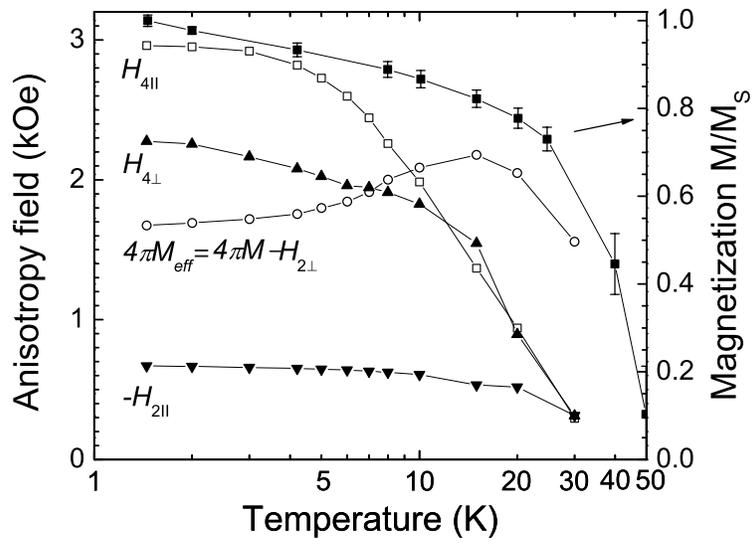

Fig. 8  X. Liu *et al.*

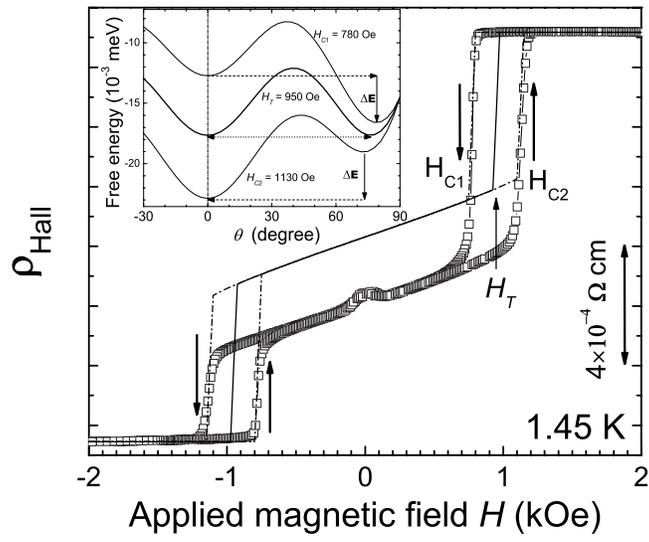

Fig. 9  X. Liu *et al.*